# How to Identify Investor's types in real financial markets by means of agent based simulation

Filippo Neri

Department of Electrical and Computer Engineering,
University of Naples, Naples
ProMarket 11 srl, Milan
email:filippo.neri.email@gmail.com

**Abstract**

The paper proposes a computational adaptation of the principles underlying principal component analysis with agent based simulation in order to produce a novel modeling methodology for financial time series and financial markets. Goal of the proposed methodology is to find a reduced set of investor's models (agents) which is able to ap- proximate or explain a target financial time series. As computational testbed for the study, we choose the learning system L-FABS which combines simulated annealing with agent based simulation for approx- imating financial time series. We will also comment on how L-FABS's architecture could exploit parallel computation to scale when dealing with massive agent simulations. Two experimental case studies show- ing the efficacy of the proposed methodology are reported.

# 1 Introduction

In this paper, we show how the principles of Principal Component Analysis [12] can be combined with Agent based Simulation and Evolutionary Computation resulting in a novel methodology for identifying investor's types in financial markets. The discussed methodology would discover a computational function whose independent variables are models of investor's behaviors and whose dependent variable is the target market financial time series. The computational function is implemented by means of software agents. By exploiting the PCA principles, we will show how to build subsets of investor's behaviors that are able to approximate the financial time series. Any subset of models could then be considered a simpler explanatory model both for the studied market.

The learning system L-FABS [17, 18], which combines simulated annealing with an agent based simulation, will be used as experimental testbed. An additional advantage of using agent based simulation as implemented in L-FABS is that, because the execution of each agent computational procedure is independent from the others, L-FABS can exploit *parallel processing* and *scale* relatively easily when large markets, i.e. made of a large number of investors, have to be analysed.

To frame the complexity of the modeling task at hand, let us consider that because the relationship among investors' models used by the machine learn- ing system L-FABS is not linear, the plain application of the PCA method- ology as used in Statistics cannot be applied. Indeed, given the iterative and intricate interactions, yet computationally definable, among several investors that make up a financial market, it would not make any sense trying to de- termine simple linear correlations among them. We will return to this point below with a specific case.

A key feature that we would like to point out and also one of the challenges of our work consists in the use of real financial time series. Our research will not use artificial time series generated by artificial financial markets because we believe that works studying artificially generated financial time series may either overfit the hypothetical time series' generative model or the artificial time series may simply miss to display important properties that can be found in real ones. That is the reason why in our work we only use publicly available and real financial time series.

When modeling data in financial time series (values of a dependent variable in Statistics' terminology) using linear regression, it is quite common to involuntarily over select many regression variables (independent variables/components) which either have poor statistical relations with the dependent variable or
that are strongly correlated with other independent variables occurring in the regression equation. Thus resulting in a linear regression equation (or model) with a redundancy of explanatory/independent variables.

Consider for instance the following multiple linear regression explaining the SP500 data by using a number of independent variables (such as the Gross Domestic Product, the Consumer Price index / inflation index, etc.):

$$SP\,500(today) = constantK + \alpha * SP\,500(yesterday) + \beta$$
$$* \,GDP\,(currentmonth) + \gamma * CPI(currentmonth) +$$
$$\delta * yields of 1 Year Treasury Bonds(today) +$$
$$\theta * rainy Days In New York(past Year)...$$

In the above regression equation, we may have that some of the variables (case 1) may have no linear regression relation with the dependent variable, like in the case hopefully of the number of rainy days in New York in the past year, or (case 2) may have some linear relations with the dependent variable but also be correlated with one or more other independent variables in the linear regression equation, like in the case of the CPI and the yield of 1 year Treasury Bonds. For case 1, the multiple linear regression expression will contain a coefficient near to zero for the variable thus signalling that the independent variable does not contribute to the explanation of the dependent variable (null/low explanatory power). For case 2, the use of principal component analysis will help identifying which independent variables make the higher contribution to the prediction of the dependent variable (high explanatory power). Thus PCA can guide the researcher in removing from the regression equation those variables/components/factors that have low explanatory power. The resulting regression equation will then contains only some of the original independent variables: those with the highest explanatory power.

In the above example, it is reasonable to estimate linear relationships among the supposedly independent variables because the regression equation is based on Econometrics' theory. However if the hypothesized model for the time series would assume the existance of non linear interactions or if the hypothesized model would be a recursive combination of time indexed variables or be procedurally defined, like in the case of agent based modeling, then it would be meaningless to test for linear correlation or for other linear relationships among the variables. Thus all the array of statistical tests used to test for linear correlation and based on the assumption of normal distribution of the variable values actually became void of utility when dealing with sophisticated and interrelated components.

The rest of the paper is organized as follows: in Section 2, we comment on how agent based modeling can be used to model financial time series; in Section 3 and 4, we describe our methodology and how the approximation error between two time series can be measured; Section 5 and 6 report the commented experimental analysis; Section 7 shows a comparison of L-FABS to other learning systems, and in Section 8, we draw our conclusions.

## 2 Agent based modeling as a computation tool for evaluating models of investors

By considering the structure of financial markets, it can be observed that any market as a whole is made up of the investment decisions of many individuals, the investors. If we try and match this structural perspective about financial markets with the observations made by researchers in the agent based modeling community, we open up the possibility of studying financial markets by using agent based computational simulation. In fact the state- of-the-art literature shows that agent based modeling is a flexible modeling methodology for simulating several types of domains [3, 4, 27, 7, 10] including consumer markets [26], economies [8] or societies [9] and financial time series [2, 11, 18]. Moreover examples of how evolutionary computation [5, 6] and agent based modeling can be used to deal with economic tasks can be found in [1, 25]. The cited papers have been selected with the only intent to provide examples of the listed approaches and without any claim to be an exhaustive list of previous work on the topic.

A classic approach to model a financial market by using agent based simulation would be to define the behavior of a group of investors as a set of decision making algorithms that could be implemented into a computational procedure (an agent). Then a simulator could be run to make several so defined agents interact the ones with the others in order to reproduce that particular behavior of the financial market that has to be studied. As already said, in this paper, we will use a machine learning systems: the Learning Financial Agent Based Simulator L-FABS [17] as the computational context where sets of investors' models can be combined to approximate a given financial time series. Because of L-FABS's modular architecture where all the agents/investors are instantiated and run independently one from the others, L-FABS would allow for an easy *parallel implementation* where several processors would run many agents' decision making procedures in parallel. Therefore L-FABS could both *exploit parallel processing* and *scale relatively easily when massive markets*, made of a large number of investors, may need to be investigated. We refer the reader to [17] for a detailed description of the L-FABS architecture, which may allow the reader to obtain a better understanding of how investor's behaviors can be modeled by in an agent based system to explain/predict financial time series. Also different research aspects of L-FABS have been studied in [23, 22, 21, 19, 18, 16].

For sake of completeness, we mention the fact that alternative approaches to model investors' or customers' behavior in a variety markets or trading situations also exists as discussed for instance in [14, 20].

## 3 Applying the PCA principles to find a re- duced set of agent based models of investors

The approach that we describe here can be used to discover simpler explanatory/predictive models of a target financial time series in terms of a computational combination of a set of investors' behaviors. As computational tool to define and manage different models of investors, we will use the system L-FABS. L-FBAS is essentially an agent based simulator com- bined with a machine learning algorithm. The machine learning algorithm is used to discover good simulation's parameters so that the simulated time se- ries can closely approximate the target one. If this were the case, the learned computation model, which would include several agent based models, could then be thought as a simplified computational representation of the financial market that has generated the target financial time series. By applying the PCA principles to the investors/agent models learned by L-FABS then a reduced set of investors' models may be found, with respect to the input one, while still mantaining an high explanatory power for the target time series. Here is how the PCA principles are adapted and employed in our methodology:

a) start with an hypothetical and possibly redundant set Orig of investor's behavior models and run Orig in L-FABS. The output agent based model produced by Orig and its approximation error will act as the benchmark model and error.
b) measure the explanatory power of each of the models in Orig
c) add to an initially empty set Reduced the individual model with the highest explanatory power in Orig, that is with the lowest approximation error. Remove the selected model from Orig.
d) repeat step c) until the approximation error of the set Reduced, when run in L-FABS, is better than or close enough to the approximation error of model Orig as built in step a).

The informed reader would have recognized in the above algorithm a classic hill climbing procedure as described in any artificial intelligence textbook. Of course we are aware of the limitations of getting stuck on local maxima when using a simple hill climbing method but the selection of the best optimization method is not the focus of the research discussed here. Here we are indeed making the point and empirically showing that PCA principles can be suc- cessfully combined with agent based simulation to discover sets of investors' models. We leave to a future work the investigation of if and how using a better optimization function could improve the composition of the discov- ered reduced set of investors' models. This future investigation would also open the door to exploring connections with meta-learning studies that have appeared in the machine learning community, just as an example reference [24].

# 4 Measuring the explanatory power of mod- els

In order to measure the explanatory power of a set of investors' behaviors, first we will codify them into a set of agents that can be run into L-FABS, then we will run L-FABS with the objective to approximate a target financial time series and, finally, we will measure the Mean Absolute Percentage Error (MAPE) of the predicted time series with respect to the target one [18].

We selected the Mean Absolute Percentage Error or MAPE function to measure the approximation error between two time series because it is commonly used in Statistics when two data samplings have to be compared.

MAPE is defined as:
$$MAPE(X, Y) = \frac{1}{N} \sum_{i=1}^{N} \left| \frac{x_i - y_i}{x_i} \right|$$

Given two time series X and Y, the lower the values for MAPE, the closer the two are. Thus the lower the MAPE value, the highest the explanatory power of the agent based model (set of investor's behaviors) run in L-FABS.

## 5  Experimental analysis

The financial time series selected for our experiments consists of a period of the SP500 index from 3 Jan 2008 to 20 Aug 2010. As usual when working with learning systems, we will train L-FABS on a part of the dataset, the learning set, and then we will use the remaining part of the dataset as test set to assess the performances of the learned model. We then divided the original period in a learning set: SP500 data from 3 Jan 2008 to 31 Dec 2008 and a test set: SP500 from 2 Jan 2009 to 20 Aug 2010.

Also, the interested reader, may note that we will run L-FABS configured in the partial knowledge (PK) operating modality. In the PK modality, only the starting point of the time series, t=0, is given to L-FABS in order to initialize the simulation. In this configuration, the time series model in L-FABS will move from one predicted value for the time series to the next without knowing/using the correct value of the time series at time t-1 in order to estimate a value for time t. We selected the PK scenario because the predicted time series will not make use of any other information apart from *the value of the target time series at time 0* and the information coded and expressed by *the set of models of investors behaviors*.

In the experimental settings, we will explore two different sets of agent based models of investor's behavior denoted as Configuration A and Configu- ration B. To keep things simple, we will use four types of investors (Financial Agents) to capture the variety of investment decisions and the variety of size of financial transactions that occur in real financial markets. According to each investor's type, many agents (Financial Agents) are then created in the simulation with similar but not identical behavior. The four types of investors that we model can be thought as: individual investors (and the likes), banks (and the likes), hedge funds (and the likes), and central banks[1]

---
[1] A little thought objection to our choice to include *Central Banks* among the actors

(and the likes). They differ in term of the size of the assets they can invest in financial markets and for their risk/reward appetite. In addition, their numerical presence is also different. As already said two different configurations of investor types (or two different set of agent based models) have been studies, we will identify them as Configuration A (Table 1) and Configuration B (Table 2).

Table 1: *Configuration A of Investor types*

| Investor type | Total Assets per investor type (in millions) | Number of |
|---|---|---|
| Individual | 0.1 | 150 |
| Funds | 100 | 100 |
| Banks | 1000 | 245 |
| Govt/Central Banks | 10000 | 5 |

Table 2: *Configuration B of Investor types*

| Investor type | Total Assets per Investor type (in millions) | Number of |
|---|---|---|
| Individual | 0.1 | 150 |
| Funds | 100 | 100 |
| Banks | 1000 | 245 |
| Govt/Central Banks | 100000 | 5 |

*Case Configuration A*
Let start by observing the performances of L-FABS when all the investors' models are used: as it can be seen in fig. 1a, L-FABS is able to output a predicted time series that is very close to the real one. The corresponding MAPE errors for all configurations are reported in Table 3.

If we disable all the models for the investors, the output of L-FABS become a constant value equal to the value of the target time series at time 0 as expected and causing a very high MAPE error as reported in fig. 1b.

---

influencing financial markets has to be easily and strongly rejected considering how, in recent decades, Central Banks have acted to 'pump up' financial markets by adding large quantities of liquidity to the related faltering real economies. Thus Central Banks, who have always acted in the background, have finally lost their image of neutral agents with respect to the financial systems.

Let us consider what happens when only one type of investors can act in approximating the financial time series, figg. 1c, 1d, 1e, and 1f. As it can be seen from the graphs, each type of investor has an explanatory power ranging from high to low when it came to model the target time series. If we also look at the MAPE values, we can observe that just by using the Institutional Investors/Banks type of investors' model we can achieve optimal approximation of the target time series. Thus all the other investors' types are redundant in this case.

Applying the adated principal component methodology, described in Sec- tion 3, to model selection in this case is then trivial: in fact it will reduce the original set of investor's types by selecting the model associated with Institutional Investors/Banks.

*Case Configuration B*

Let us then observe the results obtained when L-FABS is run in the Con- figuration B case. Again starting with all the investors' models, in fig. 2a, L-FABS is able to output a predicted time series that is very close to the tar- get one. The corresponding MAPE errors for all configurations are reported in Table 4. When only one type of investors is used, for example see fig. 2b and 2c, each type of investors displays its own explanatory power ranging from high to low when it came to model the target time series. The behavior of using only the models for Retail/Private Investors or Hedge Funds is the same as observed in Case Configuration A and so not reported given their limited expanatory power. Instead we report in fig. 2d what will happen if both of them are used to predict the target time series.

Table 3: MAPE values for Case Configuration A. The reported MAPE values are averaged on 10 runs. To keep the table readable, we report the standard deviation only for the lowest and closest values of MAPE.

| case | MAPE |
|------|------|
| a | 3.21 ± 0.04% |
| b | 13.14 ± 0.00% |
| c | 9.48 ± 0.00% |
| d | 13.05 ± 0.00% |
| e | 11.29 ± 0.00% |
| f | 3.15 ± 0.02% |

Table 4: MAPE values for Case Configuration B.

| case | MAPE |
|:----:|:----:|
| a | 3.31 ± 0.06% |
| b | 3.92 ± 0.20% |
| c | 7.14 ± 0.00% |
| d | 9.70 ± 0.00% |
| e | 3.38 ± 0.05% |

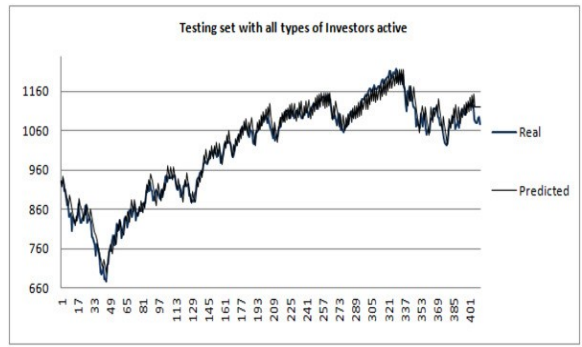

(a)

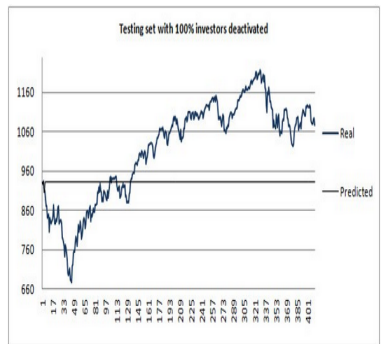

(b)

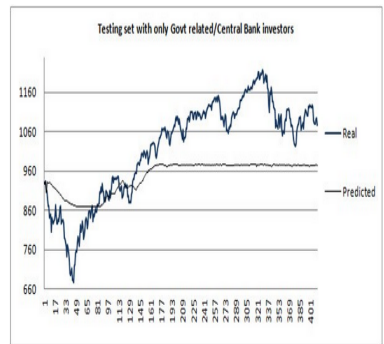

(c)

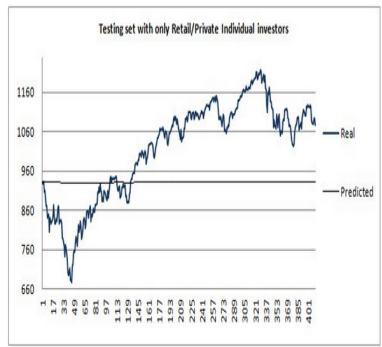

(d)

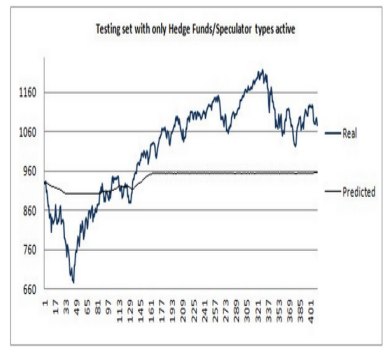

(e)

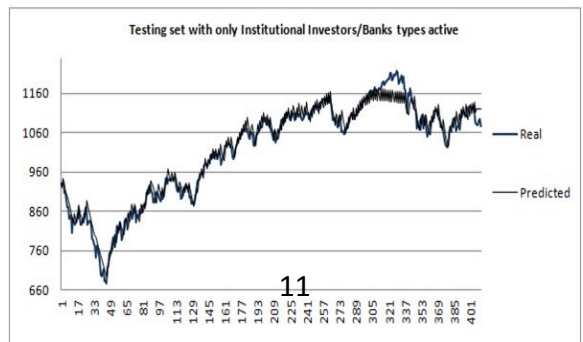

(f)

Figure 1: Comparison of the actual and predicted time series obtained by L-FABS under the experimental settings Configuration A as measured on the testing set.

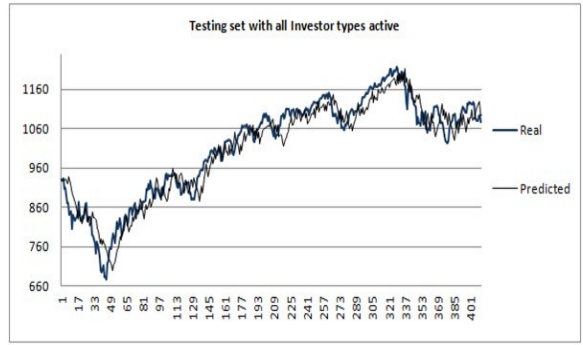
(a)

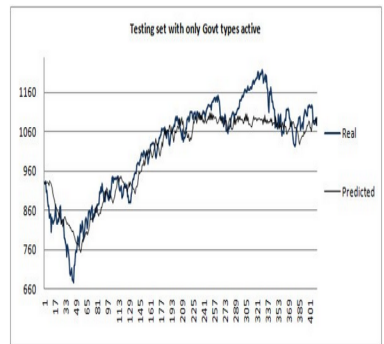
(b)

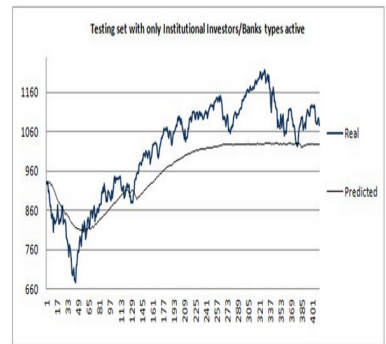
(c)

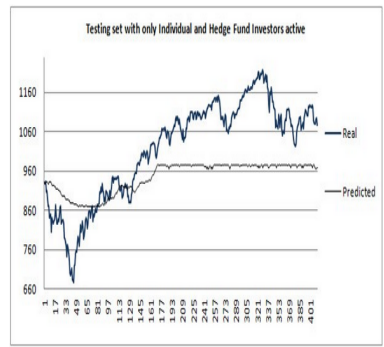
(d)

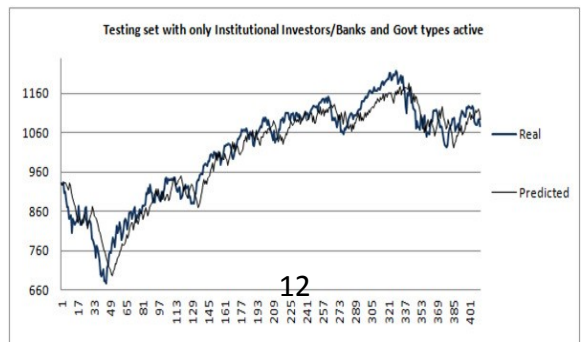
(e)

Figure 2: Comparison of the actual and predicted time series obtained by L-FABS under the experimental settings Configuration B.

The combined approximation error is better than their separate ones but still very far from the benchmark MAPE value of the original set of models. Also in Configuration B, and this is a significant difference with respect to Configuration A, none of the models for the various investors' types when taken individually is able to approximate very well the target time series. If we apply the adapted principal component methodology to this case, then we will start by selecting the model for Govt. type of investors plus the model for Institutional Investors/Banks as first candidates for a reduced set of models of types of investors' behaviors. We will thereafter run L-FABS with the two models active, and we will obtain the graph in fig. 2e and corresponding MAPE value shown in Table 4. As it can be seen the approximation result is very good and the adapted principal component methodology would stop here providing as output the reduced set of models: {Govt. investors, Institutional Investors/Banks}.

# 6 Commentary on the experiments

Note that in the reported set of experiments, we have only explored how the change in the balance of available assets among the types of investors, see the table below, can alter the composition of the reduced set of investors' models.

| Investor type | Total Assets in Configuration A (in millions) | Total Assets in Configuration B (in millions) |
|---|---|---|
| Individual | 15 | 15 |
| Funds | 10000 | 10000 |
| Banks | 245000 | 245000 |
| Govt/Central Banks | 50000 | 500000 |

The Total Assets column show that in Configuration A case, a type of investors has the majority of available assets to invest, whereas in Configuration B case the balance of available assets among investors have been changed and now there are two types of investors with similar investment capacity. However the adapted principal component methodology described here is not limited to the evaluation of only quantitative changes in a model' parameters. In fact we could use the same methodology of model simplifica- tion to explore algorithmic differences in the decision making process of the investors. Thus we could explore how different ways to participate in the market could result in more complex or simpler models of the time series. This point will be object of future research.

# 7 Experimental comparison of L-FABS to other systems

Even though a comparison of L-FABS to other learning systems is beyond the scope of the paper, we also briefly report about its performances with respect to those obtained by other learning algorithms for which enough implementation details have been reported in the literature. Results are taken from [18] and we refer to the cited paper for more information on the empirical evaluation of L-FABS on several other time series. In Table 5, we compare L-FABS, a Particle Swarm Optimization algorithm (PSO) [13], and a Multi-Layer Perceptron (MLP) [28] when operating on time series from the SP500 and DJIA respectively. These results have been reported with the aim to locate L-FABS with respect to other learning systems representative of the state of the art and without intention lay claim on the superiority of one system over the others.

Table 5: Experimental results, averaged over 10 runs, for Dataset DJIA and Dataset SP500.

| Dataset | Day to predict | PSO MAPE % | MLP MAPE % | L-FABS MAPE % |
|---------|----------------|------------|------------|---------------|
| SP500 | 1 | 0.66 | 1.00 | 0.714 ± 0.009 |
| SP500 | 7 | 1.47 | 3.11 | 1.424 ± 0.015 |
| DJIA | 1 | 0.65 | 1.06 | 0.709 ± 0.007 |
| DJIA | 7 | 1.47 | 5.64 | 1.443 ± 0.011 |

The figures in Table 5 show that the forecasting errors of L-FABS are better than those obtained by MLP and are comparable to those obtained by PSO. The figures for PSO e MLP are as reported by the authors with no confidence intervals given.

# 8 Conclusion

We proposed a computational adaptation of the principles underlying Principal Component Analysis and we implemented them in the agent based simulator L-FABS. We have also pointed out how easily and agent based simulation could allow for a parallel implementation and thus scale when markets made by a large number of investors are to be studied. The methodology we propose can be applied to the task of finding a reduced set of *models of investor's behaviors* used to approximate a target financial time series and its generative market. The reported methodology can thus be used to to evaluate the explanatory power of a set of investor's models with respect to a given market. Thus allowing also to artificially reproduce the behavior of the target market, in terms of its generated financial time series, by using L-FABS. We believe that the use of simple behavioral models, such as those found by L-FABS, can allow for a better understanding of the underlying and usually hidden mechanism that result in a macro behavior like those captured by a market index. Two case studies have been employed to show the efficacy of the proposed methodology in two instances of real financial time series. As a future research we plan to explore how to use the agent based modeling to find simplest explanatory models in new domains such as when tackling control problems [15].